# Low Temperature Crystal Structure and $^{57}$Fe Mössbauer Spectroscopy of $Sr_3Sc_2O_5Fe_2As_2$

Marcus Tegel[a], Inga Schellenberg[b], Franziska Hummel[a], Rainer Pöttgen[b] and Dirk Johrendt[a]

[a] Department Chemie und Biochemie, Ludwig-Maximilians-Universität München,
Butenandtstrasse 5–13 (Haus D), 81377 München, Germany

[b] Institut für Anorganische und Analytische Chemie, Westfälische Wilhelms-Universität Münster, Corrensstrasse 30, 48149 Münster, Germany

Reprint requests to D. Johrendt: Email johrendt@lmu.de

The crystal structure of the layered iron arsenide $Sr_3Sc_2O_5Fe_2As_2$ was determined between 300 and 10 K. The lattice parameters of the tetragonal cell decrease anisotropically according to $\frac{\Delta c}{c} : \frac{\Delta a}{a} \approx 4.2$, which results in a slight flattening of the As–Fe–As bond angle within the FeAs layers. No indication of a structural instability could be detected. $^{57}$Fe Mössbauer spectroscopic data show a single signal at 4.2, 77, and 298 K, respectively, subjected to quadrupole splitting. The isomer shift increases from 0.36(1) mm/s at 298 K to 0.49(1) mm/s at 4.2 K. No indication for magnetic ordering was found.

**Introduction**

The discovery of high-$T_c$ superconductivity in a number of iron arsenides with ZrCuSiAs [1], ThCr$_2$Si$_2$ [2] and Cu$_2$Sb type structures [3] has raised an enormous and growing interest in these class of materials [4,5]. Especially the quest for new members of this family of superconductors with even higher critical temperatures ($T_c$) attracts the attention of many research groups. Despite the substantial progress that has been made within only one year, the exact recipe that produces higher $T_c$'s is far from being clear. An empirical relation between the dimensionality of the crystal structures and the critical temperatures has been proposed. The largest $T_c$ of up to 55 K appears in the ZrCuSiAs-type compounds like Sm(O,F)FeAs [6], where the [FeAs] layers are well separated by [SmO] layers. In the ThCr$_2$Si$_2$-type materials like (Ba,K)Fe$_2$As$_2$, where the FeAs layers are only separated by barium atoms, the highest observed $T_c$ was 38 K [2] and finally in LiFeAs with even an smaller layer separation, $T_c$ decreases further down to 18 K [3]. However, the relation between $T_c$ and the separation of the FeAs layers, *i. e.* the two-dimensional character, is not justified by any theoretical argument and may have its origin in some rather artificial relationships to the cuprate superconductors. Nevertheless, as long as no other signpost is available, the search for new iron arsenide materials with low dimensional structures is a promising task. Recently, new compounds derived from structures with isoelectronic copper sulfide (CuS) layers [7-9] were reported [10,11], among them the superconductors Sr$_2$ScO$_3$FeP ($T_c$ = 17 K) [12] and Sr$_2$VO$_3$FeAs ($T_c$ = 37 K) [13]. However, these compounds become superconducting without doping and they also lack the supposed preconditions for superconductivity in iron arsenides, because they show neither structural distortions nor antiferromagnetic ordering. Thus, the origin of superconductivity remains unclear in these new compounds just as its absence in the previously reported compound Sr$_3$Sc$_2$O$_5$Fe$_2$As$_2$ [10]. In order to complete the structural and magnetic data of this compound, we present the low temperature crystal structure and a $^{57}$Fe Mössbauer spectroscopy study of (Sr$_3$Sc$_2$O$_5$)Fe$_2$As$_2$ in this paper. We also give some crystallographic relationships that may be useful in the search of new layered iron based superconductors.

**Experimental**

*Synthesis*

($Sr_3Sc_2O_5$)$Fe_2As_2$ was synthesized by heating a stoichiometric mixture of strontium, scandium, iron (II) oxide and arsenic oxide in an alumina crucible sealed in a silica ampoule under an atmosphere of purified argon. The mixture was heated to 1323 K at a rate of 200 K/h, kept at this temperature for 60 h and cooled down to room temperature. The product was homogenized in an agate mortar, pressed into a pellet and sintered at 1323 K for 60 h.

*Crystal structure determination*

Powder patterns were recorded on a Huber G670 Guinier imaging plate diffractometer (Cu-$K_{\alpha 1}$ radiation, Ge-111 monochromator) equipped with a closed-cycle He-cryostat. Rietveld refinements were performed with the TOPAS package [14] using the fundamental parameters approach as reflection profiles (convolution of appropriate source emission profiles with axial instrument contributions as well as crystallite microstructure effects). In order to describe small peak half width and shape anisotropy effects, the approach of *Le Bail* and *Jouanneaux* [15] was implemented into the TOPAS program and the according parameters were allowed to refine freely at 300 and 10 K. Preferred orientation of the crystallites was described with a March-Dollase function. An empirical $2\theta$-dependent absorption correction for the different absorption lengths of the Guinier geometry was applied. In order to get the accurate course of the lattice parameters, powder patterns between 10 and 300 K were refined using a similar approach as described in Ref. [16]. As the background between 10 and 25 degrees $2\theta$ shows artifacts from the low-temperature configuration of the Guinier diffractometer, small sections of this range were excluded from the refinements.

$^{57}$Fe *Mössbauer* spectroscopy

A $^{57}$Co/Rh source was available for the $^{57}$Fe Mössbauer spectroscopy investigations and the quoted values of the isomer shifts are given relative to this material. The (Sr$_3$Sc$_2$O$_5$)Fe$_2$As$_2$ sample was placed in a thin-walled PVC container. The measurement was run in the usual transmission geometry at temperatures between 4.2 and 298 K. The source was kept at room temperature. The total counting time was approximately 1 day per spectrum. Fitting of the spectra was performed with the NORMOS-90 program system [17].

**Results and discussion**

Crystal *chemistry*

The refined crystal structure of Sr$_3$Sc$_2$O$_5$Fe$_2$As$_2$ (Fig. 1) at room temperature is in good agreement with Ref. [10]. A detailed description can also be found in Ref. [18]. No evidence of any structural instability at low temperatures was detected. The crystallographic data of Sr$_3$Sc$_2$O$_5$Fe$_2$As$_2$ at 300 and 10 K are compiled in Table 2. The course of the lattice parameters and As–Fe–As angles on cooling reveals no anomaly as shown in Figure 3. While the *a* lattice parameter only decreases by 0.61 pm, the *c* lattice parameter decreases by 16.7 pm on cooling down to 10 K. Thus the thermal contraction of the unit cell is anisotropic according to $\frac{\Delta c}{c} : \frac{\Delta a}{a} \approx 4.2$. The more pronounced shrinkage of the *c* axis leads to a slight increase of the vertical As–Fe–As angle (ε of Fig. 1) by about 1°, *i. e.* the FeAs layers become flatter.

The structure of Sr$_3$Sc$_2$O$_5$Fe$_2$As$_2$, space group *I*4/*mmm*, Wyckoff sequence *ge*$^3$*dba* is closely related to other structures, which however have different composition. The different site occupancy variants and the corresponding free *z* parameters for the 8*g* (0, ½, *z*) and 4*e* (0, 0, *z*) sites are listed in Tables 1 and 2. The first representative of this structure type was the mineral chalcothallite (K/Tl)$_2$Cu$_7$SbS$_4$ [19]. Later on, the ternary germanides SmNi$_3$Ge$_3$ [20] and U$_3$Co$_4$Ge$_7$ [21] and the quaternary and quinary compounds listed in Table 2 have been reported.

The large structural diversity of this structure type enables formation of [Fe$_2$As$_2$], [Cu$_2$S$_2$], [Co$_2$Ge$_2$], [Ni$_2$Ge$_2$], and [Pt$_2$P$_2$] tetrahedral layers, which can be separated by either intermetallic or oxide layers, leading to different bonding patterns. The tetrahedral layers are built up from the atoms at Wyckoff positions 4$e$ (4$^{th}$ column of Table 2) and 4$d$ (6$^{th}$ column of Table 2). It is interesting to note that for some representatives the transition metal atoms switch between the 4$e$ and 4$d$ sites, but this does not scale with the course of the electronegativities.

Although the atoms occupy the same Wyckoff positions, nature allows for variance in the different compounds, *i. e.* the lattice parameters and the free $z$ parameters of the 8$g$ and 4$e$ sites. This allows large flexibility for this structural arrangement. A careful inspection of the $z$ parameters listed in Table 2 readily reveals differences between the eight compounds. Based on this comparison, we can regroup the eight compounds into two groups, (i) Sr$_3$Sc$_2$O$_5$Fe$_2$As$_2$, Sr$_3$Fe$_2$O$_5$Cu$_2$S$_2$, (Sr$_3$Sc$_2$O$_5$)Cu$_2$S$_2$, and (K/Tl)$_2$Cu$_7$SbS$_4$, and (ii) U$_3$Co$_4$Ge$_7$, SmNi$_3$Ge$_3$, Eu$_2$Pt$_7$AlP$_{2.95}$, and Eu$_2$Pt$_{7.3}$Mg$_{0.7}$P$_3$. Since these structural differences significantly affect the chemical bonding, these two groups of compounds are isopointal [22,23] rather than strictly isotypic. The short overview on these eight compounds manifests the large potential of these and other stacking variants of tetrahedral layers and one can expect rich crystal chemistry.

$^{57}$Fe Mössbauer *spectroscopy*

The $^{57}$Fe Mössbauer spectra of the Sr$_3$Sc$_2$O$_5$Fe$_2$As$_2$ sample at various temperatures are presented in Figure 4 together with transmission integral fits. The corresponding fitting parameters are listed in Table 3. In accordance with the presence of a single Fe site we observe a single signal at an isomer shift of $\delta$ = 0.36(1) mm/s and an experimental line width $\Gamma$ = 0.34(1) mm/s subject to quadrupole splitting of $\Delta E_Q$ = 0.19(1) mm/s at room temperature. The non-cubic site symmetry ($\bar{4}m2$) of the iron atoms is reflected in the quadrupole splitting value. For 77 and 4 K we observe also a single signal at an isomer shift of $\delta$ = 0.47(1) mm/s, respectively 0.49(1) mm/s. The increase of the isomer shift with decreasing temperature can be considered as a second order Doppler shift (SODS). These parameters compare well with the recently reported $^{57}$Fe data for LaFeAsO [24,25], LaFePO [26], SrFe$_2$As$_2$ [27], and BaFe$_2$As$_2$ [28]. Down to 4.2 K the $^{57}$Fe spectra of Sr$_3$Sc$_2$O$_5$Fe$_2$As$_2$ give no hint for magnetic ordering.

**Table 1.** Crystallographic data for $Sr_3Sc_2O_5Fe_2As_2$ at 300 K and 10 K.

| | | |
|---|---|---|
| temperature | 300 K | 10 K |
| space group | $I4/mmm$ | $I4/mmm$ |
| molar mass (g mol$^{-1}$) | 694.302 | 694.302 |
| lattice parameters (pm) | $a$ = 407.81(1) | $a$ = 407.20(1) |
| | $c$ = 2683.86(5) | $c$ = 2667.19(5) |
| cell volume, (nm$^3$) | 0.44635(2) | 0.44225(1) |
| density, (g cm$^{-3}$) | 5.17 | 5.214 |
| $\mu$ (mm$^{-1}$) | 68.5 | 69.1 |
| Z | 2 | 2 |
| data points | 18298 | 17575 |
| reflections | 99 | 99 |
| $d$ range | 1.008 – 13.419 | 1.006 – 13.336 |
| excluded $2\theta$ range(s) | 14.5 – 17.5 | 11.3 – 13.0; 14.0 – 18.9 |
| constraints | 1 | 1 |
| atomic variables | 10 | 10 |
| profile variables | 6 | 6 |
| anisotropy variables | 24 | 24 |
| background variables | 60 | 60 |
| other variables | 5 | 5 |
| $R_P$, $wR_P$ | 0.014, 0.018 | 0.014, 0.019 |
| $R_{bragg}$, $\chi^2$ | 0.005, 0.832 | 0.006, 0.810 |
| **Atomic parameters:** | | |
| Sr1 | $2b$ (0, 0, ½); $U_{iso}$ = 152(5) | $2b$ (0, 0, ½); $U_{iso}$ = 46(5) |
| Sr2 | $4e$ (0, 0, $z$); $z$ = 0.3604(1); $U_{iso}$ = 63(4) | $4e$ (0, 0, $z$); $z$ = 0.3601(1); $U_{iso}$ = 26(4) |
| Sc1 | $4e$ (0, 0, $z$); $z$ = 0.0727(1); $U_{iso}$ = 116(7) | $4e$ (0, 0, $z$); $z$ = 0.0730(1); $U_{iso}$ = 58(7) |
| Fe1 | $4d$ (0, ½, ¼); $U_{iso}$ = 115(5) | $4d$ (0, ½, ¼); $U_{iso}$ = 33(5) |
| As1 | $4e$ (0, 0, $z$); $z$ = 0.1996(1); $U_{iso}$ = 151(5) | $4e$ (0, 0, $z$); $z$ = 0.2002(1); $U_{iso}$ = 83(5) |
| O1 | $8g$ (0, ½, $z$); $z$ = 0.0828(1); $U_{iso}$ = 97(10) | $8g$ (0, ½, $z$); $z$ = 0.0826(1); $U_{iso}$ = 79(10) |
| O2 | $2a$ (0, 0, 0); $U_{iso}$ = 97(10) | $2a$ (0, 0, 0); $U_{iso}$ = 79(10) |
| **Selected bond lengths (pm) and angles (degrees):** | | |
| Sr – O | 254.6(2)×4; 288.4(1)×4; 301.6(3)×8 | 254.7(2)×4; 287.9(1)×4; 299.9(3)×8 |
| Sc – O | 195.1(2)×1; 205.7(1)×4 | 194.8(2)×1; 205.2(1)×4 |
| Fe – Fe | 288.4(1)×4 | 287.9(1)×4 |
| Fe – As | 244.8(1)×4 | 243.1(1)×4 |
| As – Fe – As | 107.8(1)×4; 112.8(1)×2 ($\varepsilon$) | 107.4(1)×4; 113.8(1)×2 ($\varepsilon$) |
| O – Sc – O | 89.0(1)×4; 97.6(1)×4; 164.9(2)×2 | 89.1(1)×4; 97.1(1)×4; 165.8(2) |

**Table 2.** Site occupancies for tetragonal structures with the Pearson symbol tI28 (no. 139), space group $I4/mmm$, and the Wyckoff sequence ge$^3$dba. The $z$ parameters of the 8$g$ and 4$e$ sites (if refined) are listed in parentheses. These $z$ parameters correspond to settings 0, ½, $z$ for the 8$g$ and 0, 0, $z$ for the 4$e$ sites. For details see text.

| Compound | 8$g$ ($z$) | 4$e$ ($z$) | 4$e$ ($z$) | 4$e$ ($z$) | 4$d$ | 2$b$ | 2$a$ | Ref. |
|---|---|---|---|---|---|---|---|---|
| Sr$_3$Sc$_2$O$_5$Fe$_2$As$_2$ | O1 (0.0828) | Sr2 (0.6396) | As (0.1996) | Sc (0.0727) | Fe | Sr1 | O2 | – |
| Sr$_3$Fe$_2$O$_5$Cu$_2$S$_2$ | O1 (0.0798) | Sr2 (0.6418) | S (0.1956) | Fe (0.0718) | Cu | Sr1 | O2 | [29] |
| Sr$_3$Sc$_2$O$_5$Cu$_2$S$_2$ | O1 (0.087) | Sr2 (0.6431) | S (0.195) | Sc (0.073) | Cu | Sr1 | O2 | [30] |
| U$_3$Co$_4$Ge$_7$ | Ge2 (0.08041) | U2 (0.67001) | Ge3 (0.2025) | Co1 (0.1174) | Co2 | Ge1 | U1 | [21] |
| SmNi$_3$Ge$_3$ | Ni1 (0.0553) | Sm (0.6535) | Ni2 (0.1999) | Ge1 (0.1063) | Ge2 | Ge4 | Ge3 | [20] |
| Eu$_2$Pt$_7$AlP$_{2.95}$ | Pt3 (0.0705) | Eu (0.6652) | Pt1 (0.2074) | P2 (0.1195) | P1 | Al | Pt2 | [31] |
| Eu$_2$Pt$_{7.3}$Mg$_{0.7}$P$_3$ | Pt3 (0.0734) | Eu (0.6665) | Pt1 (0.2080) | P2 (0.1214) | P1 | 0.7Mg+0.3Pt | Pt2 | [31] |
| (K/Tl)$_2$Cu$_7$SbS$_4$ | Cu1 (0.0570) | Tl+K (0.6483) | S2 (0.214) | S1 (0.093) | Cu2 | Sb | Cu3 | [19] |

**Table 3.** Fitting parameters of $^{57}$Fe Mössbauer spectra of Sr$_3$Sc$_2$O$_5$Fe$_2$As$_2$ at different temperatures. $\delta$: isomer shift; $\Gamma$: experimental line width, $\Delta E_Q$: electric quadrupole splitting parameter (for details see text).

| T / K | $\delta$ / mms$^{-1}$ | $\Gamma$ / mms$^{-1}$ | $\Delta E_Q$ / mms$^{-1}$ |
|---|---|---|---|
| 298 | 0.36(1) | 0.34(1) | 0.19(1) |
| 77 | 0.47(1) | 0.30(1) | 0.21(1) |
| 4.2 | 0.49(1) | 0.32(1) | 0.20(1) |

**Figure captions**

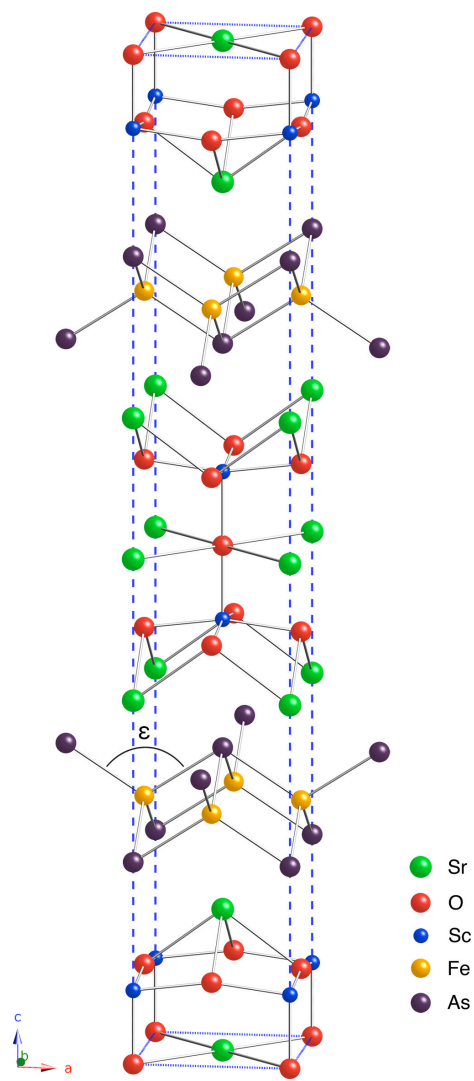

**Fig. 1**. Crystal structure of $Sr_3Sc_2O_5Fe_2As_2$.

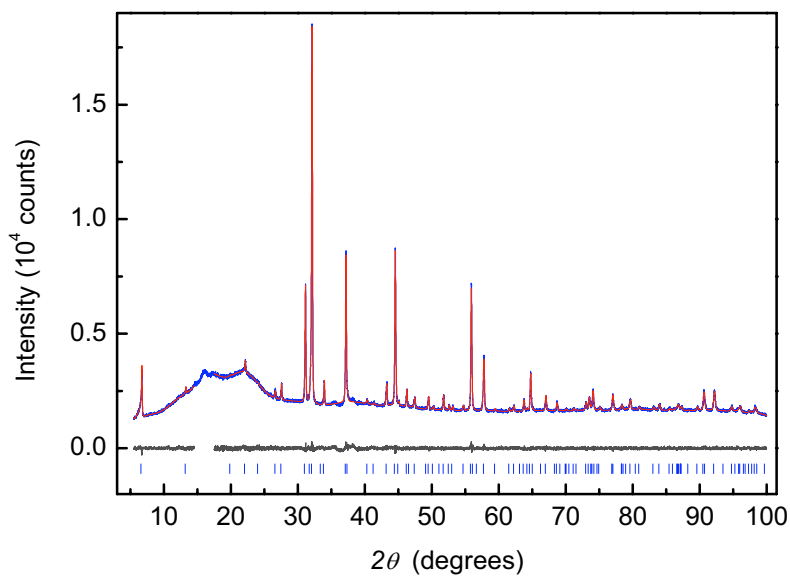

**Fig. 2**. Powder pattern (blue) and Rietveld refinement (red) of $Sr_3Sc_2O_5Fe_2As_2$ at 300 K.

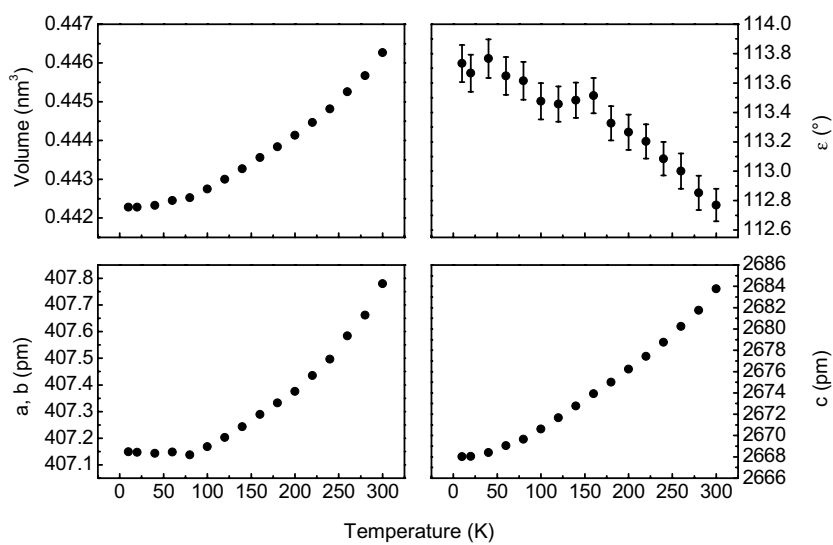

**Fig. 3**. Lattice parameters and the vertical As–Fe–As angle ($\varepsilon$) of $Sr_3Sc_2O_5Fe_2As_2$ at various temperatures. For the lattice parameters the error bars are within data points.

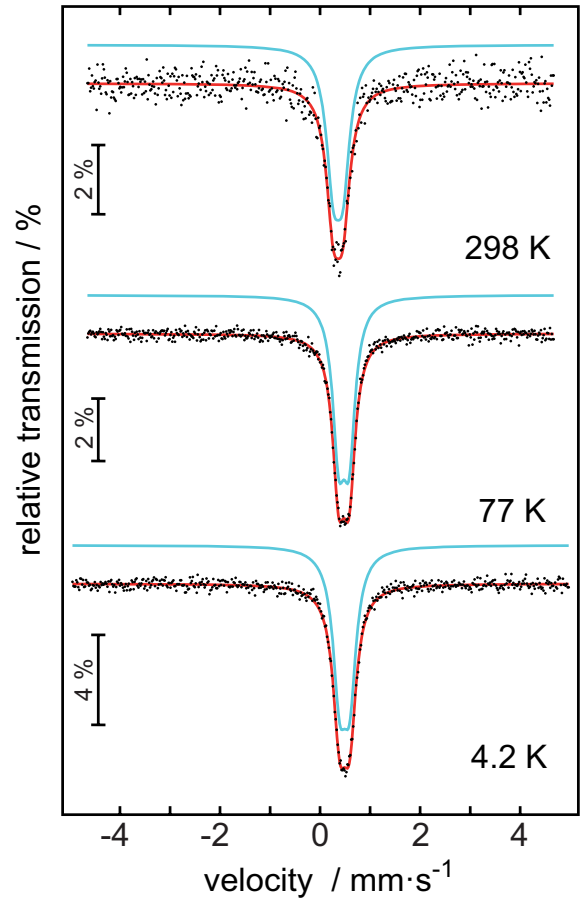

**Fig. 4**. Experimental and simulated $^{57}$Fe Mössbauer spectra of $Sr_3Sc_2O_5Fe_2As_2$ at various temperatures.